\documentclass[aps,prb,groupeaddress,supperscriptaddress,showpacs,footinbib,twocolumn,showkeys,final]{revtex4}
\usepackage{graphics,graphicx,color}
\begin{document}

\title{Low temperature dephasing in irradiated metallic wires}
\author{Thibaut~Capron$^{1}$, Yasuhiro~Niimi$^{1}$, Fran\c{c}ois~Mallet$^{1}$, Yannick~Baines$^{1}$, Dominique~Mailly$^{2}$, Fang-Yuh~Lo$^{3}$,
 Alexander~Melnikov$^{3}$, Andreas~D.~Wieck$^{3}$, Laurent~Saminadayar$^{1,\dagger}$ and Christopher~B\"auerle$^{1,\dagger}$}
\affiliation{$^{1}$Institut  N\'{e}el, CNRS and Universit\'{e}
Joseph Fourier, B.P. 166, 38042 Grenoble, France}
\affiliation{$^{2}$Laboratoire de Photonique et Nanostructures,
CNRS, route de Nozay, 91460 Marcoussis, France}
\affiliation{$^{3}$Lehrstuhl f\"{u}r Angewandte
Festk\"{o}rperphysik, Ruhr-Universit\"{a}t, Universit\"atsstra{\ss}e
150, 44780 Bochum, Germany}

\date{\today}

\begin{abstract}
We present phase coherence time measurements in
quasi-one-dimensional Ag wires implanted with Ag$^{+}$ ions with an
energy of $100\,keV$. The measurements have been carried out in the
temperature range from $100\,mK$ up to $10\,K$; this has to be
compared with the Kondo temperature of iron in silver, \textit{i.e.}
$T_{K}^{Ag/Fe} \approx 4\,K$, used in recent experiments on
dephasing in Kondo systems [F. Mallet \textsl{et al.}, Phys. Rev. Lett. \textbf{97}, 226804 (2006); G. M. Alzoubi and N. O. Birge, Phys. Rev. Lett. \textbf{97}, 226803 (2006)]. We show
that the phase coherence time is not affected by the implantation
procedure, clearly proving that ion implantation process by itself
\emph{does not lead to any extra dephasing} at low temperature.
\end{abstract}
\maketitle

The temperature dependence of the electronic phase coherence time in
mesoscopic samples has been the subject of a heavy debate in recent
years. This is due to the fact that the low temperature coherence
time is related to the lifetime of the quasiparticles in a Fermi
liquid, one of the main ingredients of Landau's Fermi Liquid theory.
As the phase space available for electron diffusion (at the Fermi
energy) crunches to zero at zero temperature, it sounds reasonable
that the phase coherence time $\tau_{\phi}$ of the electrons should
diverge as the temperature $T$ goes to zero. The frequent
experimental observation of a saturating phase coherence time at low
temperature in mesoscopic samples \cite{mohanty_prl_97} has thus
stimulated a lot of theoretical work in order to prove that it is
intrinsic \cite{GZ_prl_98} or on the contrary related to extrinsic
mechanisms \cite{imry_epl_99,zawa_prl_99,pierre_prb_03}. Among
several extrinsic mechanisms, the most obvious candidate is the
diffusion by Kondo impurities.

Diffusion by magnetic impurities has been known for a long time to
be an efficient mechanism for
decoherence\cite{saminadayar_physicaE_07}. Comparison of experiment
with theory has been difficult due to the absence of an adequate
theory. In the past, the high ($T\gg T_K$), with $T_K$ the Kondo
temperature) and very low ($T\ll T_K$) temperature limit have been
treated by theory \cite{glazmann_prb_03a+b}. Only very recently it
has been possible to compare experimental results on the decoherence
rate due to Kondo impurities with an exact theoretical calculation
over all the temperature range from zero temperature to well above
$T_{K}$. This important step has been made possible by the use of
Numerical Renormalisation Group (NRG) technics by two different
groups \cite{zarand_prl_04,rosch_prl_06,zarand_prb_07,rosch_prb_07}
and has led to a tremendous progress in the understanding of
dephasing due to Kondo impurities. Indeed, recent experimental works
\cite{schopfer_prl_03, bauerle_prl_05,mallet_prl_06,birge_prl_06}
have confirmed the main features of the NRG calculation, using the
Kondo systems Au/Fe and Ag/Fe. Due to the different Kondo
temperatures of these two systems $(T_K^{Au/Fe}\,\sim
0.5K;\,T_K^{Ag/Fe}\,\sim 4K)$, experiments could cover a temperature
range extending from $\sim5\,T_{K}$ down to $\sim0.01\,T_{K}$. In
almost the entire temperature range investigated, \emph{the
experiments have shown that the electron dephasing due to magnetic
impurities is remarkably well described by a spin $S = 1/2$, single
channel Kondo model}. This shows that magnetic impurities are
perfectly screened when the temperature is lowered well below
$T_{K}$.

Only at very low temperature $T \leq 0.1\,T_{K}$, small but
significant deviations to the perfect $S = 1/2$ single channel Kondo
model have been observed \cite{mallet_prl_06,birge_prl_06}. These
deviations have been found to be proportional to the implanted
magnetic impurity concentration. They must hence be due to the
magnetic impurities themselves, or to the implantation process. It
has been argued that they might be attributed to two levels systems
created during the implantation of high energy ions: crystal defects
would be created and some of them would be unstable. Such dynamical
defects can indeed lead to decoherence even at very low temperature.
On the other hand, it is well known that the Kondo temperature of a
given system depends on the coupling constant between the magnetic
ions and the conduction electrons; if some magnetic ions were in a
position different than a lattice site, then this may lead to a
Kondo effect with a  different (and possibly much lower) Kondo
temperature than the well known Kondo temperature of $4\,K$ of the
bulk AgFe system. As these deviations are observed in the very low
temperature regime where the Fermi liquid should be observed, it is
of fundamental importance to understand whether these are due to the
\emph{magnetic} nature of the implanted ions or to the implantation
process itself.

In this article, we present phase coherence time measurements in
silver quantum wires implanted with silver ions in a temperature
range from $10K \approx 4T_{K}^{Ag/Fe}$ down to $0.1K
\approx0.02\,T_{K}^{Ag/Fe}$. We show that no extra dephasing is
observed when compared to the dephasing rate measured in the same
but unimplanted sample. This rules out that the possibility of low
temperature dephasing by dynamical two levels systems created during
the implantation process and hence shows that the anomalous low
temperature dephasing observed in references
\cite{mallet_prl_06,birge_prl_06} is solely due to the magnetic
nature of the impurities.

\begin{figure}[thb]
\includegraphics[width=8cm]{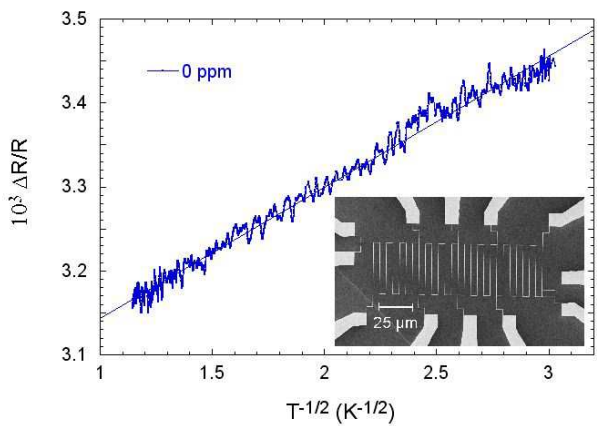}
\caption{(color online) Temperature dependence of the resistance at
a field of B=540\,mT, sufficient to suppress the weak localisation
contribution. The solid line corresponds to a fit based on
equation~\ref{eq_R_T}. The inset shows a scanning microscope
micrograph of the sample.} \label{R_T}
\end{figure}

Samples (see inset of figure \ref{R_T}) are fabricated on silicon substrate using electron beam
lithography on polymethyl-methacrylate resist. No adhesion-layer has
been used, and the silver has been evaporated in an electron gun
evaporator from a $99.9999\,\%$ purity source. The electrical and
geometrical parameters are summarized in table \ref{table}.

\begin{table}[h]
\squeezetable
\begin{tabular}{c c c c c c c c c}
{Sample}&{$l$}&{$R$}&{$\rho$}&{$D$}&{dose}&{$n_{s}$}\\
{}&{($\mu$m)}&{($\Omega$)}&{($\mu\Omega cm$)}&
{(cm$^{2}$/s)}&{($ions/cm^2$)}&{$(ppm)$}\\
\hline
\vspace{-2mm} \\
{$Ag_{Ag1}$}&{$270$}&{$1049$}&{$2.91$}&{$148$}&{$0$}&{$0$}\\
{$Ag_{Ag2}$}&{$310$}&{$1605$}&{$3.88$}&{$111$}&{$2.5\cdot 10^{13}$}&{$70$}\\
{$Ag_{Ag3}$}&{$270$}&{$1549$}&{$4.30$}&{$100$}&{$1.0\cdot 10^{13}$}&{$30$} \\
\end{tabular}
\caption{Sample characteristics: $ l, R, \rho$, $D$ and $n_{s}$
correspond to the length, electrical resistance, resistivity,
diffusion coefficient, and implanted Ag ion concentration,
respectively. All samples have a width of $w=150\,nm$ and thickness
of $t=50\,nm$. The diffusion constant has been calculated via
equation $D = 1/3\,v_{F}\,l_{e}$, with $v_{F}$ the Fermi
velocity and $l_{e}$ the mean free path.}
\label{table}
\end{table}

All samples have been fabricated in a single evaporation run. Sample
$Ag_{Ag2}$ and samples $Ag_{Ag3}$ have been ion implanted with
$Ag^{+}$ ions at an energy of $100\,keV$ and a concentration $n_{s}$
= 70\,ppm and 30\,ppm, while sample $Ag_{Ag1}$ has been left
unimplanted for reference.

Resistance measurements have been performed using a standard $ac$
lock-in technique and a very low noise preamplifier
($0.4\,nV/\sqrt{Hz}$) situated at room temperature. At low
temperature, the electrical resistance of a quasi one-dimensional
wire is given by \cite{gilles_book}:
\begin{equation}
R = R_{0} + 0.782\,\lambda_{\sigma}\,R^2/R_K\,L_T/L \label{eq.R_T}
\label{eq_R_T}
\end{equation}
where $R_{0}$ is the residual resistance, $L_T=\sqrt{\hbar D/k_B T}$
the thermal length and $R_K=h/e^2$; $\lambda_{\sigma}$ a parameter
which represents the strength of the screening of the interactions
in the metal. Resistance of the sample $Ag_{Ag1}$ as a function of
$1/\sqrt{T}$ is depicted on figure~\ref{R_T}: the experimental data
follow nicely the theoretical prediction, proving that the electrons
are indeed cooled down to the lowest temperature ($\sim100\,mK$) of
the experiment. From the fit to equation~\ref{eq.R_T} we obtain a
parameter $\lambda_{\sigma} = 3.77$, in relatively good agreement
with the theoretical value $\lambda_{\sigma}^{theo} = 3.16$ as
determined in previous measurements \cite{saminadayar_physicaE_07}.

The phase coherence time is obtained by fitting the low field
magnetoresistance to the standard weak localisation
theory\cite{hikami_80}. The magnetic field span is $\pm4000\,G$ at
temperatures above 1K and $\pm400\,G$ at low temperatures. The
spin-orbit length $L_{so}$ is determined by fitting the
magnetoresistance at high temperature and yields a value of  $L_{so}
= 330\,nm$. This value is then kept fixed for the entire fitting
procedure and $L_{\phi}$ remains the only adjustable parameter.

\begin{figure}[thb]
\begin{center}
\includegraphics[width=10cm]{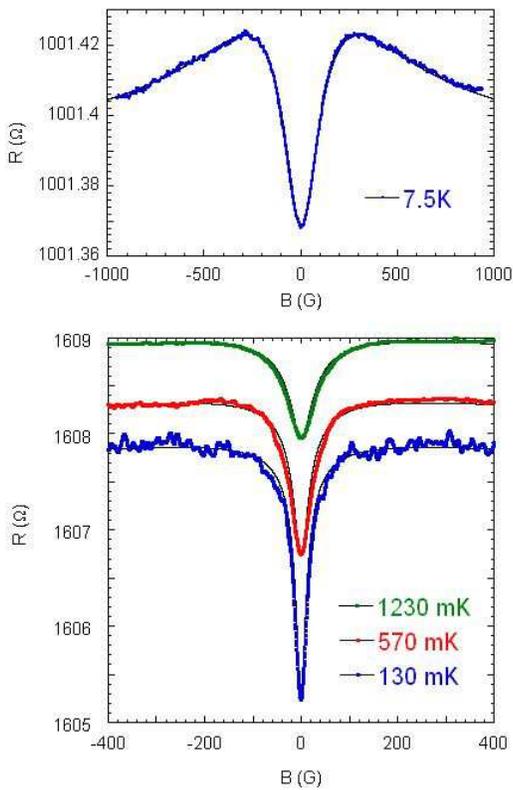}
\caption{(color online) Magnetoresistance of sample $Ag_{Ag1}$ (top)
and $Ag_{Ag2}$ (bottom) for several temperatures. The data at high
temperature (top) allows us to determine the spin-orbit length. The
solid lines are fits to the standard weak localisation theory.}
\label{magnetoR}
\end{center}
\end{figure}

\begin{figure}[thb]
\includegraphics[width=8cm]{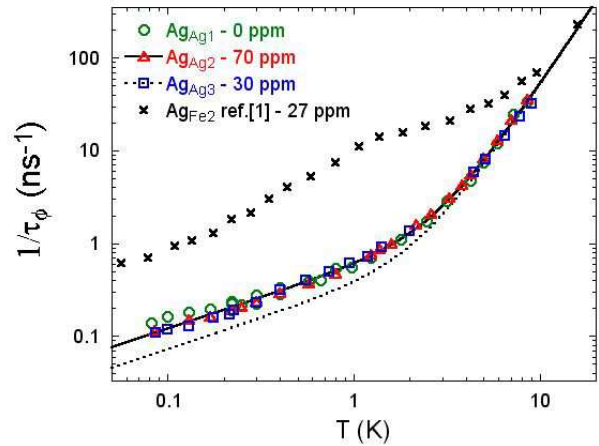}
\caption{(color online) Inverse of the phase coherence time as a
function of temperature for samples $Ag_{Ag1}$, $Ag_{Ag2}$ and
$Ag_{Ag3}$ as well as $Ag_{Fe2}$ from ref. \cite{mallet_prl_06}. The
solid line is the best fit to equation~\protect{\ref{eq_tauphi}}.
The dotted line is the theoretical expectation for sample
$Ag_{Ag1}$.} \label{tau_phi}
\end{figure}

Several magnetoresistance curves for the unimplanted as well as
implanted samples are displayed on figure~\ref{magnetoR}. Agreement
between experimental data and theoretical fits is nearly
perfect and allows us a reliable determination of the phase
coherence length $L_{\phi}$ and subsequently of the phase coherence
time \textsl{via} the relation $\tau_{\phi} = L_{\phi}^{2}/D$. For metallic
wires containing no magnetic impurities, the phase coherence time is
given by the formula:

\begin{equation}
\frac{1}{\tau_\phi} = \frac{1}{\tau_{e-e}} + \frac{1}{\tau_{e-ph}}
\label{eq_tauphi}
\end{equation}
where
\begin{equation}
\frac{1}{\tau_{e-e}}=a_{theo}\,T^{2/3}=\bigg[\frac{\pi}{\sqrt{2}}\frac{R}{R_K}\frac{k_B}{\hbar}\frac{\sqrt{D}}{L}\bigg]^{2/3}T^{2/3}
\label{alpha}
\end{equation}
corresponds to the Altshuler-Aronov-Khmelnitsky (AAK) expression for
the electron-electron interaction term\cite{AAK_82,gilles_book} and
$1/\tau_{e-ph} = b\,T^{3}$ to the electron-phonon interaction term.
From the fit of our data for the unimplanted sample, we obtain the
experimental curve $\tau_{\phi}(T)$ depicted on figure~\ref{tau_phi}
and experimental coefficients $a_{exp} = 0.56\,ns^{-1}K^{-2/3}$ and
$b = 0.05\,ns^{-1}K^{-3}$. This has to be compared with the
theoretical value $a_{theo} = 0.30\,ns^{-1}K^{-2/3}$. The agreement
between the two values is quite good, although the experimentally
observed dephasing time is obviously lower than the theoretically
expected one. It must be stressed that this is quite a general
feature of phase coherence time measurements obtained from weak
localisation: even when the theoretical power law ($T^{2/3}$) is
observed, the prefactor is always larger than the one expected from
the standard AAK theory \cite{saminadayar_physicaE_07}.

The same analysis can be made for the samples implanted with silver
ions. The phase coherence time as a function of temperature is
displayed on figure~\ref{tau_phi}: within the error bars of our
measurements, the experimental data follow nicely the theoretical
expression of equation~\ref{eq_tauphi} with the same prefactors
$a_{exp}$ and $b$. More importantly, \emph{the phase coherence time
measured in the samples implanted with high energy silver ions has
exactly the same value as in the unimplanted one}.

What can we infer from these findings? In very recent experimental
works on dephasing in Kondo
systems\cite{mallet_prl_06,birge_prl_06}, it has been shown that the
electronic phase coherence rate due to magnetic impurities
 $\gamma_{m}$  is very well described by the $S = 1/2$, single
channel Kondo model. In particular, the behavior of $\gamma_{m}$ in
a temperature range from above $T_{K}$ down to $0.1\,T_{K}$ is
extremely well described by the NRG calculations: this means that
the physics of the screening process of the magnetic impurities by
the surrounding conduction electrons is well captured by the
theoretical model. At very low temperature, typically below
$0.1\,T_{K}$, small but significant deviations from the theoretical
predictions have been observed; these deviations are of importance,
since they appear exactly in the temperature range for which one
should recover the Fermi liquid behavior of the Kondo system.
Different scenarii have been evoked to explain this anomalous
behavior at very low temperature: first, it has been suggested that
the disorder may lead to a distribution of the Kondo temperature
over a certain temperature range\cite{kettemann_prb_07}, and as a
consequence certain impurities may still be unscreened even well
below $T/{T_{K}^{Ag/Fe}}$. Although plausible, the \textquotedblleft
weak\textquotedblright~disorder present in our samples should lead
to a very narrow distribution of the Kondo temperatures and
therefore cannot account for the observed behavior.

Another and very plausible hypothesis is the creation of dynamical
defects during the ion implantation process. Ion implantation is a
relatively \textit{violent} non-equilibrium process. During the
random diffusion of the implanted ions, they create many defects
along their trajectories inside the crystal. For instance, some
silver atoms may be pushed away from their crystalline sites and may
\textquotedblleft oscillate\textquotedblright~between two unstable
positions. Such a dynamical two-levels systems (TLS) may be the
origin of a very efficient mechanism for decoherence. As the
characteristic energies of these two-levels systems should be widely
distributed, they would lead to decoherence in a large temperature
range.

\begin{figure}[thb]
\includegraphics[width=9cm]{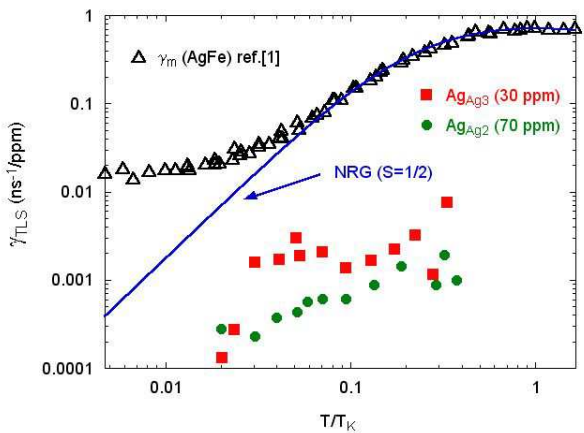}
\caption{(color online) Dephasing rate per implanted ion
concentration of hypothetical TLS as a function of temperature. For
comparison we also plot the magnetic dephasing rate of reference
\protect{\cite{mallet_prl_06}}. The solid line corresponds to the
NRG data for spin 1/2 of reference \protect{\cite{rosch_prl_06}}}
\label{tau_tls}
\end{figure}

To show that this scenario can be excluded, we represent in figure
\ref{tau_tls} the maximal contribution to dephasing of these
hypothetical TLS. If there is an additional contribution to
dephasing due to dynamical TLS, we can account for this by adding an
additional term $(1/\tau_{TLS})$ in equation (2). In order to
determine $\gamma_{TLS}\equiv(1/\tau_{TLS})$ we subtract the
electron-electron contribution as well as the electron-phonon
contribution, as indicated by the solid line in figure 3, exactly in
the same way as has been done in reference \cite{mallet_prl_06} to
obtain the magnetic decoherence rate. On the same figure we also
plot the magnetic dephasing rate from recent experiments on AgFe
quantum wires \cite{mallet_prl_06}, as well as the NRG calculations
for the spin 1/2, single channel model \cite{rosch_prl_06}.

The maximum dephasing due to dynamical TLS lie more than an order of
magnitude below the depasing rate $\gamma_m$ obtained when Fe ions
are implanted rather than Ag ions. In addition $\gamma_{TLS}$ does
not scale with the number of implanted Ag ions. It is therefore
clear that our experiment definitely rules out this scenario and that the deviations
from theoretical predictions observed in recent
works can only originate from the magnetic nature of the implanted impurities.

In order to explain the experimentally observed deviations from the
perfectly screened Kondo model one has to infer that a small
fraction of the magnetic ions remains partially unscreened, even at
the lowest temperature. The most probable scenario is that some of
the implanted ions end up at a different lattice site or in an
interstitial site inside the silver crystal compared to most of the
implanted ions. The magnetic coupling with the conduction electrons
of these magnetic ions would be different, and should lead to a much
lower Kondo temperature. It is thus highly desirable to confirm this
scenario by ab-initio calculations. Actually, from the magnetic
decoherence rate observed at very low temperature, it can be
estimated that a fraction of only $2\,\%$ of the impurities in such
unconventional sites would be enough to produce the observed
decoherence rate at very low temperature: this hypothesis sounds
reasonable and may indeed explain the experimental data.

Finally we would like to point out that in a very recent theory a
free magnetic moment phase is predicted for low dimensional
disordered conductors \cite{kettemann_07}. When the magnetic
coupling is small, local wave function correlations can leave some
spins paramagnetic even at the lowest temperatures, and a finite
dephasing rate at zero temperature would be expected. A quantitative
analysis of such a scenario is in progress.

In conclusion we have measured the phase coherence time of silver
quantum wires implanted with high energy silver ions. The phase
coherence times measured in both the implanted and non-implanted
wires exhibit the same temperature
dependence, clearly proving that the implantation process by
itself does not lead to any additional dephasing.

\acknowledgments We are indebted to the Quantronics group  for the
use of its evaporator and silver source. We acknowledge helpful
discussions with A. Rosch, T.~A.~Costi, S. Kettemann, D. Feinberg,
P. Simon, S. Florens, L. L\'evy, G. Zar\'and, L. Borda and A.
Zawadowski. This work has been supported by the European Comission
FP6 NMP-3 project 505457-1 \textquotedblleft Ultra
1D\textquotedblright and the \textsl{Agence Nationale de la
Recherche} under the grant PNANO \textquotedblleft
QuSpin\textquotedblright. L.S acknowledges financial support from
the \textsl{\textit{Institut Universitaire de France}}.


\begin{thebibliography}{99}
\bibitem[$\dagger$]{mail}{Mail to: \texttt{QuSpin@grenoble.cnrs.fr}}
\bibitem{mallet_prl_06} F. Mallet \textsl{et al.}, Phys. Rev. Lett. \textbf{97}, 226804 (2006).
\bibitem{birge_prl_06}G. M. Alzoubi and N. O. Birge, Phys. Rev. Lett. \textbf{97}, 226803 (2006).
\bibitem{mohanty_prl_97} P. Mohanty, E.M.Q Jariwala, and R.A. Webb, Phys. Rev. Lett. {\bf 78}, 3366 (1997).
\bibitem{GZ_prl_98}D. S. Golubev and A. D. Zaikin, Phys. Rev. Lett. {\bf 81}, 1074 (1998).
\bibitem{imry_epl_99} Y. Imry, H. Fukuyama and P. Schwab  \textsl{et al.}, Europhys. Lett. {\bf 47}, 608 (1999).
\bibitem{zawa_prl_99} A. Zawadowski, J. v. Delft, and D. C. Ralph,  Phys. Rev. Lett. {\bf 83}, 2632 (1999).
\bibitem{pierre_prb_03} F. Pierre \textsl{et al.}, Phys. Rev. B {\bf 68}, 085413 (2003).
\bibitem{saminadayar_physicaE_07} L. Saminadayar \textsl{et al.}, Physica E {\bf 40}, 12
(2007).
\bibitem{glazmann_prb_03a+b} M.G. Vavilov, L.I. Glazman, Phys. Rev. B \textbf{67}, 115310
(2003) and M. G. Vavilov, L. I. Glazman, A. I. Larkin, Phys. Rev. B
\textbf{68}, 075119 (2003).
\bibitem{schopfer_prl_03}  F. Schopfer, C. B\"auerle, W. Rabaud, and L. Saminadayar, Phys. Rev. Lett {\bf 90}, 056801 (2003) and Adv. Solid. State Phys. {\bf 43}, 181 (2003).
\bibitem{bauerle_prl_05} C. B\"auerle, F. Mallet, F. Schopfer, D. Mailly, G. Eska, and L. Saminadayar, Phys. Rev. Lett {\bf 95}, 266805 (2005).
\bibitem{zarand_prl_04} G. Zar\'and, L. Borda, J.v. Delft, and N. Andrei, Phys. Rev. Lett. {\bf 93}, 107204 (2004).
\bibitem{rosch_prl_06} T. Micklitz, T. A. Costi, A. Altland, and A. Rosch, Phys. Rev. Lett. {\bf 96}, 226601 (2006).
\bibitem{zarand_prb_07} L. Borda \textsl{et al.}, Phys. Rev. B {\bf 75}, 235112 (2007).
\bibitem{rosch_prb_07} T. Micklitz \textsl{et al.}, Phys. Rev. B {\bf 75}, 054406 (2007).
\bibitem{gilles_book} \'{E}. Akkermans and G. Montambaux, in \textit{Mesoscopic physics of electrons and photons}, Cambridge University Press, Cambridge (2007).
\bibitem{AAK_82} B.L. Altshuler, A.G. Aronov, and D.E. Khmelnitzki, J. Phys. C {\bf 15}, 7367 (1982).
\bibitem{hikami_80} S. Hikami, A.I. Larkin, and Y. Nagaoka, Prog. Theor. Phys. {\bf 63}, 707 (1980).
\bibitem{kettemann_prb_07} S. Kettemann and E. R. Mucciolo, Phys. Rev. B {\bf 75}, 184407 (2007).
\bibitem{kettemann_07} A. Zhuravlev \textsl{et al.}, cond-mat/arXiv:0706.3456.
\end{thebibliography}
\end{document}